\documentclass{article}
\usepackage{amsmath,graphicx,mlspconf}
\usepackage{xcolor}

\usepackage[table]{xcolor}

\usepackage{adjustbox}
\usepackage{amsmath,amssymb,amsfonts}
\usepackage{algorithmic}
\usepackage{textcomp}

\usepackage{lineno}
\usepackage{enumitem}
\usepackage{bm}
\usepackage{subcaption}
\usepackage[]{color-edits}
\addauthor{ma}{red}
\addauthor{ms}{blue}
\usepackage{float}
\usepackage{fixltx2e}
\usepackage{mathrsfs}
\usepackage{algorithm}
\usepackage{hyperref} 
\usepackage{multirow}
\usepackage[skip=7pt]{caption}
\usepackage{makecell}
\usepackage{diagbox}
\usepackage{caption} 
\usepackage{wrapfig}   
\newcommand{\ie}{i.e.\ }

\usepackage{cite}
\usepackage{cleveref}
\usepackage{booktabs}
\usepackage{array}
\newcolumntype{C}[1]{>{\centering\arraybackslash}p{#1}}
\crefname{figure}{Fig.}{Figs.}
\Crefname{equation}{Eq.}{Eqs.}
\Crefname{table}{Tab.}{Tabs.}
\definecolor{customgreen}{RGB}{0,176,80}

\definecolor{mildyellow}{RGB}{255, 245, 180} 
\definecolor{custom_velvet}{RGB}{148,0,64}
\definecolor{custom_green}{RGB}{0,148,64}
\hypersetup{
    colorlinks=true,
    citecolor=custom_velvet,
    linkcolor=custom_green,
    urlcolor=custom_velvet
}

\copyrightnotice{979-8-3195-0884-3/26/\$31.00 {\copyright}2026 IEEE}

\toappear{2026 IEEE International Workshop on Machine Learning for Signal Processing, Sep.\ 28-- Oct.\ 1, 2026, Atlanta, USA}

\title{UMPIRE-Net: Unrolled Magnitude-Phase Regularization Network for Accelerated MRI }

\name{Mahdi Saberi, Toygan Kili\c c, and Mehmet Ak\c cakaya
}\address{Department of Electrical and Computer Engineering, University of Minnesota, Minneapolis, MN, USA\\
Center for Magnetic Resonance Research, University of Minnesota, Minneapolis, MN, USA}

\begin{document}

\maketitle

\begin{abstract}
MRI reconstruction from undersampled k-space measurements is an ill-posed inverse problem. Physics-driven deep learning (PD-DL) methods have shown strong performance for this task by combining the MRI forward model with learned image regularization within algorithm-unrolling frameworks. However, most existing PD-DL methods reconstruct complex-valued images directly, thereby implicitly coupling magnitude and phase within a single learned representation. This coupled regularization may be suboptimal in reconstruction settings where accurate phase modeling plays an important role, such as partial Fourier (PF) imaging, where recovery of the omitted asymmetric k-space measurements depends on the underlying image phase. In such scenarios, explicit modeling of magnitude and phase as separate components may reduce the reliance on externally estimated or predefined phase information. To this end, we propose UMPIRE-Net (Unrolled Magnitude-Phase In REgularization Network), a PD-DL method that introduces separate learned regularizers for magnitude and phase components, together with a novel data-fidelity formulation that enforces measurements consistency. We evaluate UMPIRE-Net for accelerated MRI with PF across different datasets and acceleration factors. Experimental results demonstrate that our proposed method improves reconstruction quality compared with a conventional complex-valued PD-DL baseline, yielding sharper images and reduced artifacts. Code available at: \url{https://github.com/MahdiSaberii/UMPIRE-Net}
\end{abstract}

\begin{keywords}
Phase regularization, magnitude regularization, data fidelity, partial Fourier, unrolled network
\end{keywords}

\newcommand{\cem}[1]{\textcolor{blue}{cem: #1}}
\section{Introduction}
\label{sec:intro}
Magnetic resonance imaging (MRI) is a powerful diagnostic imaging modality, but its relatively slow acquisition may lead to patient discomfort and motion artifacts~\cite{zaitsev2015motion, ghaffari2025functional,ghaffari2025connectome}. Accelerated MRI techniques address this limitation by acquiring undersampled k-space measurements, and performing image reconstruction by solving an ill-posed inverse problem~\cite{aggarwal2019MoDL,ghanaatian2026neural}. Parallel imaging (PI) leverages multi-coil sensitivity information~\cite{sense}, while compressed sensing (CS) uses image sparsity through regularized reconstruction~\cite{Lustig2007}. Partial Fourier (PF) imaging provides additional acceleration in Cartesian MRI by acquiring only a fraction of k-space along the phase-encoding direction; however, recovering the omitted measurements is challenging due to the complex-valued nature of MRI data and its dependence on image phase \cite{kim2017loraks}. Recently, deep learning (DL) methods have achieved state-of-the-art performance for accelerated MRI reconstruction~\cite{hammernik2018learning, Knoll2020ro, saberi2026training}. In particular, physics-driven deep learning (PD-DL) methods integrate learned image priors with the MRI forward encoding model through algorithm unrolling, enabling reconstruction of complex-valued images from undersampled measurements~\cite{hammernik2018learning,aggarwal2019MoDL}.

\begin{figure*}[t]
  \centering
  \includegraphics[width=0.9\linewidth]{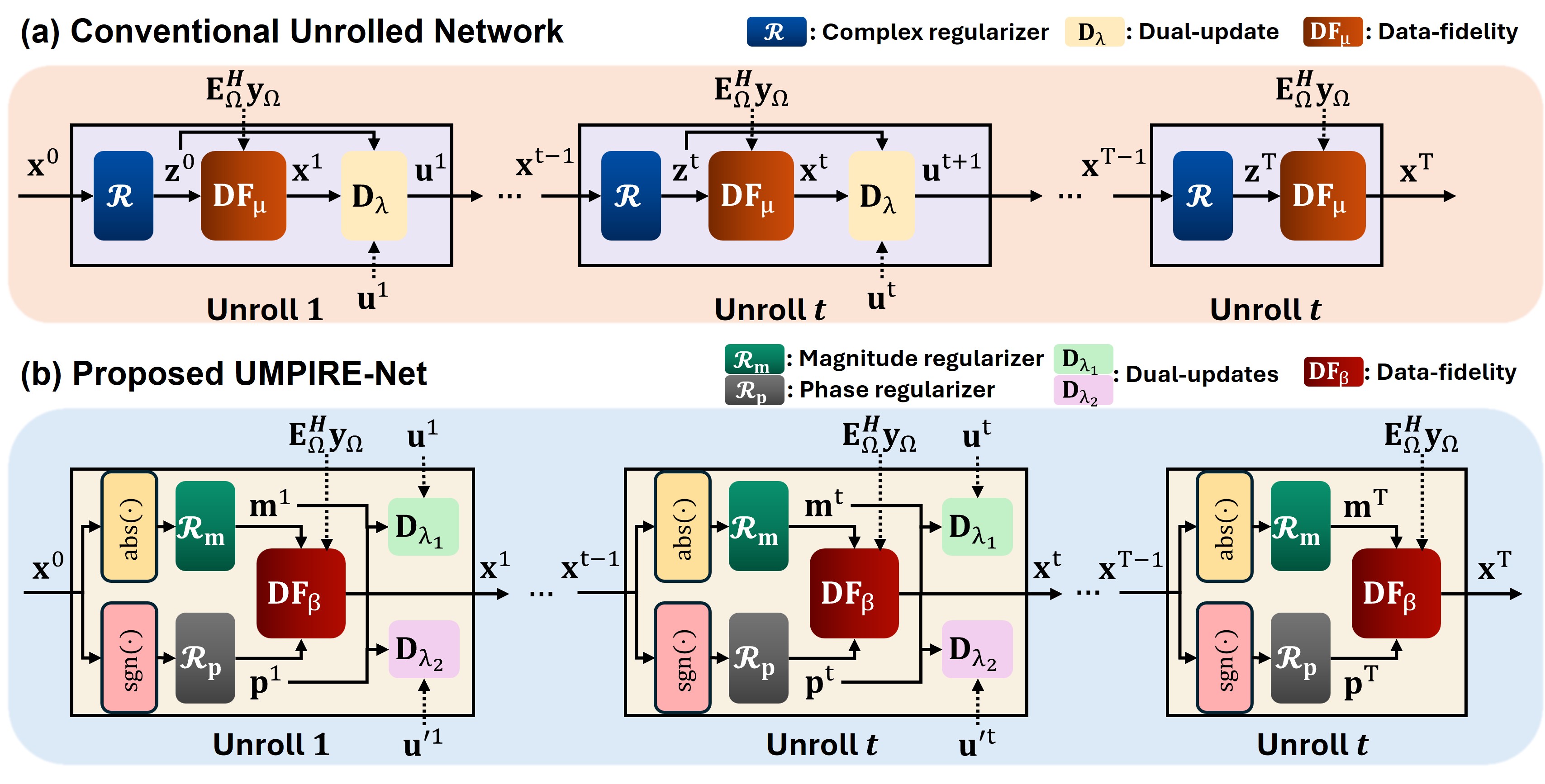}
  \caption{Comparison between conventional PD-DL and the UMPIRE-Net architecture for accelerated MRI reconstruction.}
  \label{Fig:Pipline}
\end{figure*}
Conventional PD-DL methods typically reconstruct complex images directly, thereby coupling magnitude and phase information within a single learned regularizer. Although this strategy has been highly effective for accelerated MRI, directly applying conventional PD-DL to partial Fourier data without PF-specific phase modeling or post-processing may be suboptimal. 
Since the recovery of the omitted k-space measurements in PF depends on the image phase~\cite{noll1991homodyne}, this causes issues especially for unsupervised training setups~\cite{akcakaya2022_SPMsurvey,solgi2025activation, madinei2026interlace}, where the non-acquired PF region is not directly included in the training loss, providing limited guidance for recovering these omitted k-space samples. 
On the other hand, classical PF reconstruction methods, such as homodyne reconstruction and POCS, explicitly use phase information to estimate the missing asymmetric k-space region~\cite{noll1991homodyne, mcgibney1993quantitative}. This suggests that magnitude and phase, which exhibit distinct spatial characteristics and play different roles in PF recovery, may benefit from being modeled separately within a PD-DL reconstruction framework. 

Several prior works have explored separate magnitude and phase reconstruction~\cite{zhao2012separate,ong2018general,lee2018deep}. A separate magnitude and phase reconstruction was proposed in~\cite{zhao2012separate} within a CS framework, using dedicated phase regularization to improve stability. Another line of CS-based work modeled magnitude and phase separately, and introduced phase cycling to address phase wrapping~\cite{ong2018general}. More recently, learning-based methods have used neural networks to reconstruct or refine magnitude and phase components separately~\cite{lee2018deep}. However, these approaches either rely on conventional non-DL optimization frameworks~\cite{zhao2012separate,ong2018general} or use neural networks outside a physics-driven unrolled reconstruction framework~\cite{lee2018deep}.

In this work, we propose {U}nrolled {M}agnitude-{P}hase {I}n {RE}gularization Network (UMPIRE-Net) for MRI Reconstruction. In particular, we solve the reconstruction inverse problem by independently regularizing magnitude and phase components with a novel data-fidelity unit. 
We evaluate UMPIRE-Net on accelerated Cartesian acquisitions with PF and in-plane undersampling across multiple datasets and acceleration rates. Results demonstrate that explicitly decoupling magnitude and phase regularization improves reconstruction quality compared to conventional complex-valued PD-DL methods, both quantitatively and qualitatively.

\section{Background}
\label{sec:background}
\begin{figure*}[t]
  \centering
  \includegraphics[width=0.9\textwidth]{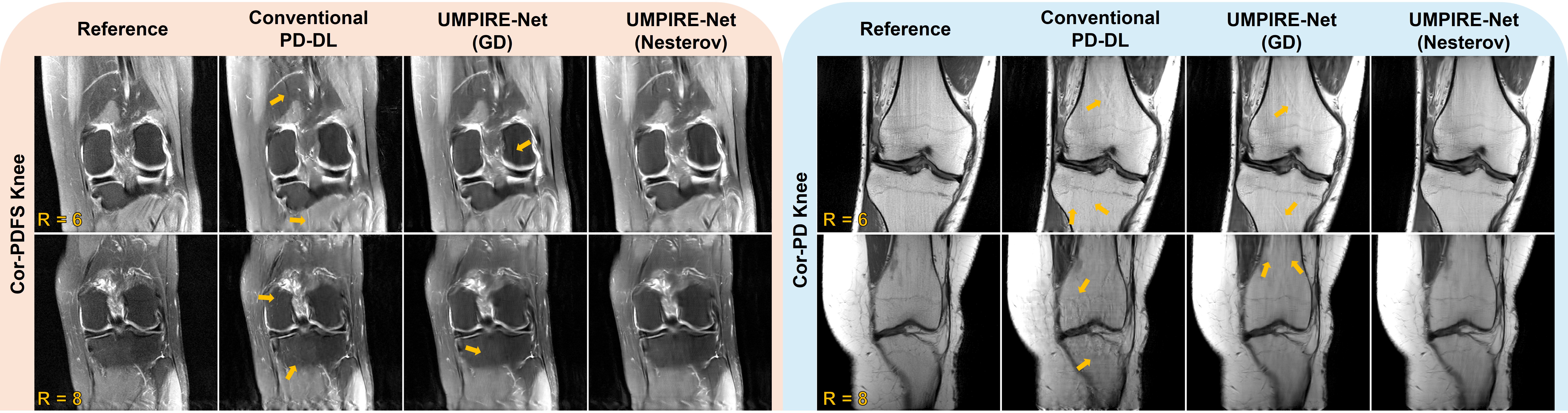}
  \caption{Representative reconstruction results for the Cor-PD and Cor-PDFS datasets. In both cases, conventional PD-DL introduces residual artifacts, indicated by the yellow arrows, which are partially reduced by UMPIRE with simple GD. These artifacts are fully removed when using UMPIRE with Nesterov-accelerated GD as the DF solver.}
  \label{Fig:pd_pdfs}
\end{figure*}
The forward model in MRI maps the image ${\bf x}$ to sub-sampled k-space measurements ${\bf y}_\Omega$ as: 
\begin{equation}
     \mathbf{y}_{\Omega} = \mathbf{E}_{\Omega} \mathbf{x} + \mathbf{n}
\end{equation}
where $\Omega$ is the undersampling pattern, $\mathbf{E}_\Omega$ is the forward multi-coil encoding operator~\cite{sense}, and $\mathbf{n}$ is complex measurement noise. The inverse problem for MRI reconstruction is generally formulated as:
\begin{equation}\label{eq:reg_least_square}
    \arg\min_{{\mathbf{x}}}{\|\mathbf{y}_{\Omega} - \mathbf{E}_{\Omega} \mathbf{x} \|_2^2 + \mathcal{R(\mathbf{x})}}
\end{equation}
where the first term enforces data fidelity (DF), and $\mathcal{R(\cdot)}$ denotes a regularizer. This regularized least square problem is traditionally solved with iterative algorithms~\cite{opt_method_fessler}, which alternates between DF and the proximal operator for $\mathcal{R(\cdot)}$. PD-DL methods unroll such iterative algorithms for a fixed number of iterations~\cite{Knoll2020ro}. Then DF is implemented using conventional methods with learnable parameters, while the proximal operator is implemented implicitly via neural networks~\cite{hammernik2018learning,aggarwal2019MoDL}. Finally, the unrolled network is trained end-to-end in a supervised or self-supervised manner~\cite{yaman2022mmssdu}. 

In this work, for conventional PD-DL methods, we unroll \Cref{eq:reg_least_square} using alternating direction method of multipliers (ADMM)\cite{saberi2026_ISBI_CMag}:
\begin{align}
&\mathbf{z}^{\mathrm{k}} 
    = \arg \min_{\mathbf{z}} \;
    \mu \|\mathbf{x}^{\mathrm{k}-1} - \mathbf{z} + \mathbf{u}^{\mathrm{k}}\|_2^2 
    + \mathcal{R}(\mathbf{z})
    \label{eq:z_update} \\ 
&\mathbf{x}^{\mathrm{k}} 
    = \arg \min_{\mathbf{x}} \;
    \|\mathbf{y}_\Omega - \mathbf{E}_\Omega \mathbf{x}\|_2^2 + \mu \|\mathbf{x} - \mathbf{z}^{\mathrm{k}} + \mathbf{u}^{\mathrm{k}}\|_2^2
    \label{eq:x_update} \\       
&\mathbf{u}^{\mathrm{k+1}} 
    = \mathbf{u}^{\mathrm{k}} 
    + \lambda(\mathbf{x}^{\mathrm{k}} - \mathbf{z}^{\mathrm{k}})
    \label{eq:u_update}
\end{align}
where $\mu$ and $\lambda$ are the learnable parameters, $\mathbf{u^{1}}=\mathbf{0}$, and $\mathbf{x^{0}} ={\bf E}_\Omega^H \mathbf{y}_\Omega$. 
Furthermore, \Cref{eq:z_update} is typically implemented with a neural network with shared weights across unrolled iterations~\cite{aggarwal2019MoDL}, while  \Cref{eq:x_update} has a closed-form solution as:
\begin{equation}\label{eq:DC_closeform}
    \mathbf{x}^{\mathrm{k}} = \big(\mathbf{E}_\Omega^H \mathbf{E}_\Omega + \mu \mathbf{I}\big)^{-1} \big({\bf E}_\Omega^H \mathbf{y}_\Omega + \mu( \mathbf{z}^{\mathrm{k}}-\mathbf{u}^{\mathrm{k}})\big)
\end{equation}
which is typically solved 
using conjugate gradient (CG)~\cite{saberi2024EMBC,akcakaya2026physics,aggarwal2019MoDL}, which itself is unrolled for a fixed number of iterations.  

\section{Proposed Method}
We reformulate the MRI reconstruction inverse problem by independently regularizing the magnitude and sign components of the image as:
\begin{equation}
    \arg \min_{{\mathbf{x}}}{\|\mathbf{y}_{\Omega} - \mathbf{E}_{\Omega} \mathbf{x} \|_2^2 + \mathcal{R}_m\big(|\mathbf{x}|\big)} + \mathcal{R}_p\left(\frac{\mathbf{x}}{|\mathbf{x}|}\right)
\end{equation}
where $\mathcal{R}_m(\cdot)$ is the magnitude regularizer, and $\mathcal{R}_p(\cdot)$ denotes the regularizer for the sign of the image. Note this is consistent with prior CS-based works~\cite{zhao2012separate}, as direct regularization of the phase leads to a non-convex objective, and the periodic nature of the phase introduces undesirable local minima~\cite{zhao2012separate}. for notational simplicity, we use $\text{sgn}(\mathbf{x}) = \frac{\mathbf{x}}{|\mathbf{x}|}$.  
\begin{align}
\label{eq:m_update}
\mathbf{m}^{\mathrm{k}} &= \arg \min_{\mathbf{m}} \beta_1 \||\mathbf{x}^{\mathrm{k}-1}|-{\bf m}+\mathbf{u}^\mathrm{k} \|_2^2 + \mathcal{R}_m(\mathbf{m})\\ 
\label{eq:p_update}
\mathbf{p}^{\mathrm{k}} &= \arg \min_{\mathbf{p}} \beta_2\| \text{sgn}(\mathbf{x}^{\mathrm{k}-1}) - \mathbf{p} + \mathbf{u'}^\mathrm{k}\|_2^2 + \mathcal{R}_p({\bf p})\\
\label{eq:x_update_umpire}
\mathbf{x}^{\mathrm{k}} &= \arg \min_{\mathbf{x}} \|\mathbf{y}_{\Omega} - \mathbf{E}_{\Omega} \mathbf{x}\|_2^2 + \beta_1 \||\mathbf{x}|-{\bf m}^{\mathrm{k}} +\mathbf{u}^\mathrm{k}\|_2^2 \nonumber \\& +\beta_2\| \text{sgn}(\mathbf{x}) - \mathbf{p}^{\mathrm{k}}+\mathbf{u'}^\mathrm{k}\|_2^2 \\
\mathbf{u}^{\mathrm{k+1}} &= \mathbf{u}^{\mathrm{k}} + \lambda_1 (|\mathbf{x}^{\mathrm{k}}| - \mathbf{m}^{\mathrm{k}})\\
\mathbf{u'}^{\mathrm{k+1}} &= \mathbf{u'}^{\mathrm{k}} + \lambda_2 (\text{sgn}(\mathbf{x}^{\mathrm{k}}) - \mathbf{p}^{\mathrm{k}})
\end{align}
\Cref{eq:m_update} and \Cref{eq:p_update} are implicitly implemented using neural networks, while the DF update in~\Cref{eq:x_update_umpire} is solved iteratively itself. To this end, following CR-calculus~\cite{kreutz2009complex}, the derivative of the objective function in~\Cref{eq:x_update_umpire} is derived with respect to $\bar{\mathbf{x}}$. However, due to the $|\cdot|$ operator, this objective is not differentiable as given. Thus, we replace $|\mathbf{x}|$ with a quadratic smoothing approximation:
\begin{equation}\label{eq:DF_smooth}
    |\mathbf{x}| \approx \big(\mathbf{x}\odot\mathbf{x}^H +\mathbf{\epsilon}\big)^{1/2} 
\end{equation}
with $\mathbf{\epsilon}>0$, and $\odot$ being the Hadamard product. Additional smoothing operators, including Log-Exp smoothing~\cite{nesterov2005smooth} and Huber smoothing~\cite{huber1964robust} are further studied in \Cref{tab:ablation}. We define
\begin{align}
    \mathcal{J}^{\mathrm{k}}(\mathbf{x}) &\triangleq \|\mathbf{y}_{\Omega} - \mathbf{E}_{\Omega} \mathbf{x}\|_2^2\\ & + \beta_1 \Big\|\big(\mathbf{x}\odot\mathbf{x}^H +\mathbf{\epsilon}\big)^{1/2} -{\bf m}^{\mathrm{k}}+\mathbf{u}^\mathrm{k} \Big\|_2^2 \nonumber \\& +\beta_2\Big\| \frac{\mathbf{x}}{(\mathbf{x}\odot\mathbf{x}^H +\mathbf{\epsilon})^{1/2}} - \mathbf{p}^{\mathrm{k}}+\mathbf{u'}^\mathrm{k}\Big\|_2^2  
\end{align}

\begin{table*}[!t]
\centering
\caption{Magnitude- and phase-related quantitative comparison of conventional PD-DL and UMPIRE for partial Fourier MM-SSDU reconstruction on Cor-PD and Cor-PDFS at {\unboldmath $\mathrm{R}\in\{6,8\}$} with PF$=6/8$. Performance gaps are reported relative to conventional PD-DL. Mean magnitude PSNR/SSIM and absolute phase error are reported, with the best result for each metric highlighted in \textbf{bold}.}
\label{tab:pf_pdfs_results}
\begin{adjustbox}{max width=\linewidth}
\renewcommand{\arraystretch}{1.1}
\setlength{\tabcolsep}{4pt}

\begin{tabular}{@{}lcccccccccccc@{}}
\toprule
\multirow{3}{*}{\large Method}
& \multicolumn{6}{c}{\large Cor-PD}
& \multicolumn{6}{c}{\large Cor-PDFS} \\
\cmidrule(lr){2-7}
\cmidrule(lr){8-13}

& \multicolumn{3}{c}{$\mathrm{R}=6$}
& \multicolumn{3}{c}{$\mathrm{R}=8$}
& \multicolumn{3}{c}{$\mathrm{R}=6$}
& \multicolumn{3}{c}{$\mathrm{R}=8$} \\
\cmidrule(lr){2-4}
\cmidrule(lr){5-7}
\cmidrule(lr){8-10}
\cmidrule(lr){11-13}

& \makecell{PSNR\\(dB)}
& \makecell{SSIM\\$(-)$}
& \makecell{$|\Delta\phi|$\\$(^\circ)$}
& \makecell{PSNR\\(dB)}
& \makecell{SSIM\\$(-)$}
& \makecell{$|\Delta\phi|$\\$(^\circ)$}
& \makecell{PSNR\\(dB)}
& \makecell{SSIM\\$(-)$}
& \makecell{$|\Delta\phi|$\\$(^\circ)$}
& \makecell{PSNR\\(dB)}
& \makecell{SSIM\\$(-)$}
& \makecell{$|\Delta\phi|$\\$(^\circ)$}\\
\midrule

Conv. PD-DL (CG)
& 35.55 & 0.914 & 19.80
& 33.64 & 0.893 & 20.96
& 32.54 & 0.750 & 30.80
& 31.47 & 0.729 & 31.57 \\
\midrule

UMPIRE (GD)
& 36.42 & 0.914 & 16.60
& 35.11 & 0.896 & 16.90
& 32.69 & 0.763 & 27.55
& 32.18 & 0.746 & 28.79 \\
\midrule

UMPIRE (Nesterov)
& \makecell{\textbf{36.76}\\
  (1.21\textcolor{customgreen}{$\uparrow$})}
& \makecell{\textbf{0.919}\\
  (.005\textcolor{customgreen}{$\uparrow$})}
& \makecell{\textbf{15.93}\\
  (3.87\textcolor{customgreen}{$\downarrow$})}

& \makecell{\textbf{35.41}\\
  (1.77\textcolor{customgreen}{$\uparrow$})}
& \makecell{\textbf{0.902}\\
  (.009\textcolor{customgreen}{$\uparrow$})}
& \makecell{\textbf{16.82}\\
  (4.14\textcolor{customgreen}{$\downarrow$})}

& \makecell{\textbf{32.83}\\
  (0.29\textcolor{customgreen}{$\uparrow$})}
& \makecell{\textbf{0.768}\\
  (.018\textcolor{customgreen}{$\uparrow$})}
& \makecell{\textbf{26.78}\\
  (4.02\textcolor{customgreen}{$\downarrow$})}

& \makecell{\textbf{32.26}\\
  (0.79\textcolor{customgreen}{$\uparrow$})}
& \makecell{\textbf{0.763}\\
  (.034\textcolor{customgreen}{$\uparrow$})}
& \makecell{\textbf{27.43}\\
  (4.14\textcolor{customgreen}{$\downarrow$})} \\
\bottomrule
\end{tabular}
\end{adjustbox}
\end{table*}

with the derivative:
\begin{align}
\nabla_{\bar{\mathbf{x}}}&\mathcal{J}^{\mathrm{k}}(\mathbf{x}) = \mathbf{E}_{\Omega}^H\big(\mathbf{E}_\Omega\mathbf{x} - \mathbf{y}_\Omega\big) \nonumber
+ \beta_1\Big(
 \frac{(\mathbf{-m}+\mathbf{u}) \odot\mathbf{x}}{(\mathbf{x}\odot\mathbf{x}^H +\mathbf{\epsilon})^{1/2} } 
 + \mathbf{x}\Big)  \nonumber\\
&+ \frac{\beta_2}{2}\Big(
\frac{\mathbf{-p}+\mathbf{u'}}{(\mathbf{x}\odot\mathbf{x}^H +\mathbf{\epsilon})^{1/2} }
 + \frac{\mathbf{x}\odot\mathbf{x}\odot (\mathbf{p}-\mathbf{u'})^H}{({\bf x}\odot{\bf x}^H +{\bf \epsilon})^{3/2}}\Big)
\end{align}
This can be used for a simple gradient descent (GD) as:
\begin{equation}\label{eq:simpleGD}
    \mathbf{x}_{\mathrm{j}}^{\mathrm{k}} = \mathbf{x}_{\mathrm{j-1}}^{\mathrm{k}} - \xi_{\mathrm{j}}^{\mathrm{k}} \nabla_{\bar{\mathbf{x}}}\mathcal{J}^{\mathrm{k}}(\mathbf{x})\Big|_{{\bf x} = \mathbf{x}_{\mathrm{j-1}}^{\mathrm{k}}} 
\end{equation}
where $\xi_{\mathrm{j}}^{\mathrm{k}}$ denotes the learnable step at $\mathrm{j}^\textrm{th}$ GD iteration of the $\mathrm{k}^\textrm{th}$ unrolled step. However, the magnitude and phase regularization terms make the optimization problem non-convex. As a result, directly optimizing \Cref{eq:x_update_umpire} with GD may converge to suboptimal local minima. Therefore, incorporating momentum-based methods may improve convergence within the DF subproblem. To this end, we use Nesterov acceleration~\cite{nesterov1983method}, which has been applied to ADMM-based optimization problems~\cite{thorley2021nesterov}, for the DF update:
\begin{align}
    \mathbf{v}_{\mathrm{j}}^{\mathrm{k}}
    &= \gamma \mathbf{v}_{\mathrm{j}-1}^{\mathrm{k}}
    + \nabla_{\bar{\mathbf{x}}}\mathcal{J}^{\mathrm{k}}(\mathbf{x})
    \Big|_{\mathbf{x}=\mathbf{x}_{\mathrm{j}-1}^{\mathrm{k}}-\gamma \mathbf{v}_{\mathrm{j}-1}^{\mathrm{k}}},
    \label{eq:Nesterov_momentum}\\
    \mathbf{x}_{\mathrm{j}}^{\mathrm{k}}
    &= \mathbf{x}_{\mathrm{j}-1}^{\mathrm{k}}
    - \xi_{\mathrm{j}}^{\mathrm{k}}\mathbf{v}_{\mathrm{j}}^{\mathrm{k}},
    \label{eq:Nesterov_update}
\end{align}
where $\mathbf{v}_{\mathrm{j}}^{\mathrm{k}}$ denotes the velocity vector with $\mathbf{v}_{0}^{\mathrm{k}}=\mathbf{0}$, and $\gamma$ is the momentum coefficient.
The overall schematic for UMPIRE-Net, along with a comparison to conventional complex-valued PD-DL, are shown in \Cref{Fig:Pipline}.

\section{Implementation Details}
\noindent \textbf{Dataset.} Experiments were performed on publicly available fully-sampled multi-coil coronal proton density (Cor-PD) and coronal proton density with fat saturation (Cor-PDFS) knee data from fastMRI database \cite{knoll2020fastmri_dataset-journal}. 
Retrospective equispaced undersampling was applied at acceleration rates $\mathrm{R} \in \{6,8\}$ to the fully-sampled data with 24 central auto-calibrated signal (ACS) lines. Partial Fourier acquisition is also applied along the phase-encoding dimension with a PF factor of $6/8$. Both knee datasets consisted of 300 slices from 10 distinct subjects for training, along with 392 slices for testing.  Coil sensitivity maps were generated from 24$\times$24 center of k-space.

\noindent \textbf{Network Architecture.} The proposed UMPIRE-Net was unrolled for ${T}=10$ iterations. The proximal steps for both the magnitude and sign images were implemented using the recently proposed time-embedded U-Net (TE-UNet) architecture~\cite{junno2025_TE-MRI_NIPS}, with $\{32, 64, 96\}$ channels in the encoder and symmetric decoding layers. The only architectural difference between the two proximal operators was the number of input channels: the magnitude proximal operator used a single input channel, since the magnitude image is real-valued, whereas the sign-image proximal operator used two input channels to represent its complex-valued input. To enforce the physical constraints of the proposed decomposition, a ReLU activation was applied to the output of the magnitude regularizer to ensure non-negativity, while the sign component was obtained from the output of the phase regularizer. The DF unit was solved using both simple GD and Nesterov-accelerated GD as described in \Cref{eq:Nesterov_update}, which itself was unrolled for 10 iterations. All the DF parameters, including $\gamma$, $\beta_1$, $\beta_2$, the step sizes $\xi_{\mathrm{j}}^\mathrm{k}$, and the dual-variable coefficients, $\lambda_1$ and $\lambda_2$ were learned and unshared across the unrolls. The smoothing constant in~\Cref{eq:DF_smooth} was chosen as $\epsilon = 10^{-4}\cdot \| \mathbf{E}_\Omega^\mathrm{H}\mathbf{y}_\Omega\|_\infty^2$.
The conventional PD-DL used a modified variant of the original TE-UNet in \cite{junno2025_TE-MRI_NIPS}, with $\{32, 64, 160\}$ channels in the encoder and symmetric decoding layers, together with a similar ADMM-based unrolling framework to ensure a fair comparison. 
Furthermore, the DF unit solved the linear objective in \Cref{eq:x_update} using 10 conjugate-gradient (CG) iterations, which converges faster than GD~\cite{aggarwal2019MoDL}. 
Thus, the main distinctions between the conventional PD-DL and UMPIRE-Net were the underlying inverse problem formulation and the corresponding DF units.

\noindent \textbf{Training.} Multi-mask self-supervised (MMSSDU) training~\cite{yaman2022mmssdu} was performed for retrospectively undersampled data using three different masks for each slice. The loss-mask ratio was set to $\rho = |\Lambda|/|\Omega| = 0.4$, where $\Theta$ and $\Lambda$ denote the disjoint training and loss mask subsets, respectively, following~\cite{yaman2022mmssdu}. The partial Fourier region was kept unsampled in the training masks (\ie, $\Theta$), allowing the networks to infer the missing k-space samples in these regions directly. A normalized $\ell_1$-$\ell_2$ loss function~\cite{saberi2026phase} was optimized using Adam with a learning rate of \(5\times 10^{-4}\). All models were trained for 100 epochs. 10\% of the training data was reserved for validation, hyperparameter fine-tuning, and learning rate scheduling. 

\begin{table}[!b]
\caption{Ablation study of smoothing operators for UMPIRE (GD) on Cor-PD at $\mathrm{R=8}$ with PF$=6/8$.}
\label{tab:ablation}
\centering
\renewcommand{\arraystretch}{1.2}
\setlength{\tabcolsep}{12.5pt}

\begin{tabular}{@{}lcc@{}}
\toprule
Smoothing Operator & \makecell{PSNR\\(dB)} &  \makecell{SSIM\\$(-)$}\\
\midrule

Log-Exp~\cite{nesterov2005smooth}
& 32.53 & 0.880 \\
\midrule

Huber~\cite{huber1964robust}
& 34.63 & 0.886 \\
\midrule

Quadratic Smoothing
& \textbf{35.11} & \textbf{0.896}\\

\bottomrule
\end{tabular}

\end{table}

\section{Results}

\Cref{tab:pf_pdfs_results} summarizes the reconstruction results at ~$\mathrm{R}\in\{6,8\}$ with PF$=6/8$ imaging on the Cor-PD and Cor-PDFS datasets. Magnitude reconstruction quality is evaluated using PSNR and SSIM. Phase accuracy is quantified by the mean absolute wrapped phase error, $|\Delta\phi|$, reported in degrees, where $\Delta\phi=\angle(\widehat{\mathbf{x}}\odot\mathbf{x}_{\mathrm{ref}}^{*})$. Across both datasets and acceleration rates, UMPIRE with Nesterov-accelerated GD achieved the best quantitative performance. This behavior is consistent with the design of UMPIRE-Net, where the magnitude and sign components are modeled separately. During training and inference, the non-acquired PF region is excluded from the sampling mask and treated as missing k-space, which requires the network to infer the missing PF k-space samples during reconstruction, allowing the proposed formulation to better utilize phase-related information than conventional PD-DL. The improvements on Cor-PDFS were milder than those observed on Cor-PD, due to the lower SNR and noisier phase characteristics of the Cor-PDFS data. These factors make phase estimation harder, reducing the relative benefit of the proposed phase-aware formulation; nevertheless, UMPIRE-Net consistently outperformed conventional PD-DL across all settings. These quantitative improvements are reflected in the representative reconstructions shown in \Cref{Fig:pd_pdfs}. Conventional PD-DL exhibits residual artifacts, highlighted by the yellow arrows, which are partially mitigated by UMPIRE with simple GD and effectively eliminated by UMPIRE-Net with Nesterov-accelerated GD.

\Cref{tab:ablation} reports the effect of different smoothing operators in UMPIRE (GD). For the additional smoothing operators, hyperparameters were tuned on a small validation subset, with $\delta=0.01$ for Huber, 
and $\alpha=20$ for Log-Exp. Despite its simplicity, the results indicate that quadratic smoothing provides a more effective approximation of $|\mathbf{x}|$ near zero.

\section{Discussion}
In this work, we proposed UMPIRE-Net, a PD-DL framework that explicitly decouples magnitude and phase regularization. Unlike conventional PD-DL approaches that directly regularize complex-valued images with a single learned prior, UMPIRE-Net introduces separate neural network proximal operators for the magnitude and phase-related components within an ADMM-based unrolled optimization framework. This formulation is motivated by  reconstruction scenarios, in which phase plays an important role, including partial Fourier acquisition as a representative application. 
The experimental results demonstrate that the proposed formulation improves reconstruction quality compared with conventional complex-valued PD-DL baselines across different datasets and acceleration rates for partial Fourier acquisition. Future work will evaluate UMPIRE-Net in applications with stronger, spatially varying phase behavior, where conventional complex-valued regularization may be further limited.

\section{Acknowledgments}
This work was partially supported by NIH R01HL179616, NIH R01EB032830, and NIH P41EB027061.

\bibliographystyle{IEEEbib}
\bibliography{refs}

\end{document}